# Microwave field vector detector based on the nonresonant spin rectification effect

Peiwen Luo, Bin Peng, Wanli Zhang and Wenxu Zhang[a]

*National key laboratory of electronic thin films and integrated devices, University of Electronic Science and Technology of China, Chengdu 611731, China*

**ABSTRACT**

Normal microwave (MW) electromagnetic field detectors convert microwave power into voltages, which results in the loss of the vector characteristics of the microwave field. In this work, we developed a MW magnetic field ($h$-field) vector detector based on the nonresonant spin rectification effect. By measuring and analyzing the angle dependence of the rectification voltages under nonresonant conditions, we can extract the three components of the $h$-field. As an initial test of this method, we obtained the $h$-field distributions at 5.4 GHz generated by a coplanar waveguide with sub-wavelength resolution. Compared to methods using ferromagnetic resonance, this technique offers a faster and more convenient way to determine the spatial distribution of the $h$-field, which can be used for MW integrated circuit optimization and fault diagnosis.

Imaging and detecting the microwave (MW) field distribution is a highly relevant element in the engineering of the MW integrated circuits, as well as for applications in atomic and solid-state physics. There are several well-established techniques to detect the MW field, one of the early methods is based on the Faraday's law. However, it is quite a challenge to scale down the detecting sensors because its sensitivity scales down with the area of the loop. In order to meet the needs of high-precision measurement of MW field at smaller scale, people have developed several other approaches such as diamond microscope,[1-3] magnetic tunnel junctions,[4,5] and scanning near field microscopy[6] to measure the MW electric ($e$-field) or magnetic field ($h$-field). Most of these techniques however lack the ability to characterize the vector characteristics of the MW field. Only recently, the vector magnetometer based on the diamond NV centers was developed.[7,8] Apart from this, methods based on the spin rectification effect (SRE) at ferromagnetic resonance, which is an effective tool in studying dynamic magnetization[9,10] and spin injection[11,12] of magnetic multilayers under MW radiation, provide another possibility to the $h$-field vector detection.[13] However, this method is restricted to the long measurement time induced by the field tuning and operation in external magnetic field intensity ($H$) on the order of several hundreds of Oester dependent on the MW frequency. Because of this, it is difficult to be applied in the high resolution and large area imagination.

The method based on the nonresonant SRE offers a promising alternative. The nonresonant SRE refers to the generation of a rectification voltage through the coupling of the $h$-field induced oscillating resistance and the $e$-field induced oscillating current

[a] Electronic addresses: xwzhang@uestc.edu.cn

($j$) at a very weak or even zero $H$.[14, 15] According to the theory proposed by Zhu. et al,[14] the ferromagnetic resonance (FMR) is not a necessary condition for SRE as the oscillation resistance can also appear at the nonresonant condition. The dynamic resistance of a magnetic layer under the superposition of a static $H$ and a MW magnetic field $\tilde{h} = he^{i\omega t}$ can be expanded as a Taylor series $\tilde{R}(\tilde{H} = H + \tilde{h}) \approx R(H) + he^{i\omega t} \frac{d\tilde{R}}{d\tilde{H}}\big|_{\tilde{H}=H}$, where the higher order terms are omitted. Obviously, there will be a non-zero oscillating resistance at the nonresonant condition as long as $dR/dH \neq 0$. They have studied the nonresonant SRE of a permalloy stripe and demonstrated that the magnitude and lineshape of SRE voltage changes with the phase difference ($\varphi$) between the $j$ and the $h$-field, while for the nonresonant signal at $H$ less than the saturation field ($H_s$) of magnetoresistance, only the magnitude changes with $\varphi$. We in this work extend the study of the nonresonant SRE to the $h$ vector detecting. Our proof-of-concept imaging experiments were performed on a millimeter MW stripline, yielding a sub-wavelength spatial resolution of the $h$-field.

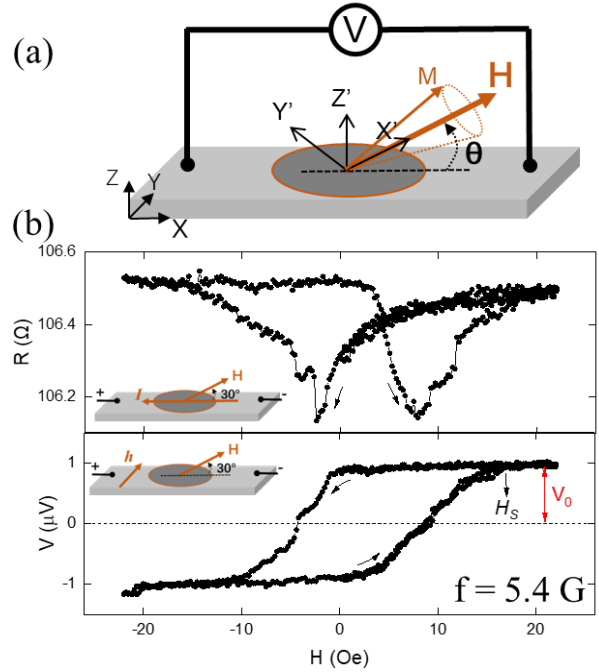

Fig. 1 (a) the coordinate system of the rectification effect measurement. (b) the magnetoresistance and nonresonant rectification voltage of a Py thin film generated with MW frequency $f$ = 5.4 GHz and $\theta$ = 30°.

Our permalloy (Py) thin film stripe is defined to be 160 μm-long and 40 μm-wide by optical lithography. The thickness of the film is about 20 nm grown on the $SiO_2$ substrate via RF magnetron sputtering at the base pressure of $2\times10^{-5}$ Pa. The coordinate system of the measurement is schematically shown in Fig. 1(a), in which the $H$ is applied in the XY plane with the angle $\theta$ to the OX direction and the voltage is measured along longitude of the stripe (X axis). We define another orthogonal coordinate system (x', y', z') for simplifying the subsequent description of the rectification effect, where the x' axis is aligned with the $H$ and the z' axis lies normal to the film plane. The magnetoresistance and nonresonant SRE of the Py film are respectively measured by a Keithley 2400 source meter and a lock-in amplifier (SR830). Fig. 1(b) shows the resistance as a function of $H$ at $\theta = 30°$ with a DC current (100 μA) flows along the X axis, the arrows show the directions of which the field varies. The resistance change is about 0.5 Ω with the anisotropic magnetoresistance (AMR) $\Delta R/R_0 \sim 0.37\%$. This AMR value is relatively low due to the fact that the sample is not annealed according to our protocols. The two dips of the butterfly like AMR curve at $\sim\pm 5$ Oe correspond to the magnetization switching of the Py stripe. Coincidentally, the voltage spectra under 5.4 GHz MW radiation shows a hysteresis loop which is consistent with the AMR curve. For the Py stripe in this work, the $dR/dH$ of the stripe is far from zero when the static field $H$ is smaller than the saturation field $H_s$ (~15 Oe), the $dR/dH$ induced rectification voltage changes with the derivations proportional to the resistance and the history of the field is therefore recorded both in the AMR and the rectification voltage curves. The consistency between them implies that the nonresonant technique is capable in electrically detecting magnetization switching processes.[16, 17]

When the magnetic field is larger than the saturation field $H_s$, the resistance of the stripe and the rectification voltage are independent of $H$, suggesting that there is no magnetization rotation as the magnetic moments are aligned in the direction of the applied field when the thin film is saturated. In this case, the magnetic moment under MW radiation will then precess consistently about the direction of the applied field. To calculate this nonresonant rectification voltage, we start from the resonant rectification theory in which the coupling between the oscillating resistance $R(t)$ and $j(t)$ is found to be:[13, 18]

$$V_0 = \frac{R_A}{M_0} sin2\theta \langle Re(j(t)) \cdot Re(m_{y'}) \rangle, \quad (1)$$

where the $R_A$ represent the AMR induced resistance change, $M_0$ is the saturation magnetization, the dynamic magnetization $m_{y'}$ of the Py stripe is related to the $h$-field by the susceptibility $\chi$ through $m_{y'} = \chi_{xx}h_{y'} + i\chi_{xy}h_{z'}$ with $\chi_{xx,xy} = (D + iL)A_{xx,xy}$, here the $\chi_{xx}$ and $\chi_{xy}$ represent the tensor components of the susceptibility, $A_{xx,xy}$ are amplitudes related to the properties of the sample, the resonant feature of the rectification voltage include both the Lorentz and anti-Lorentz lineshape determined respectively from the $D$ and $L$ term. However, for the measurement at nonresonant condition at $H \ll H_0$, the $D$ and $L$ are nearly a constant independent of $H$ as the two terms can be approximated as:

$$D = \frac{\Delta H(H-H_0)}{(H-H_0)^2+\Delta H^2} \approx \frac{-H_0 \Delta H}{H_0^2+\Delta H^2}, \quad L = \frac{\Delta H^2}{(H-H_0)^2+\Delta H^2} \approx \frac{\Delta H^2}{H_0^2+\Delta H^2}, \quad (2)$$

Where the $H_0$ and $\Delta H$ are the FMR magnetic field and linewidth. By adopting the above approximation, the Eq. (1) can be further written as:

$$V_0 = \frac{jR_A}{2M_0}\left\{\left(A_{xy}h_{z'}^r + A_{xx}h_{y'}^i\right)L + \left(A_{xy}h_{z'}^i - A_{xx}h_{y'}^r\right)D\right\}sin2\theta, \quad (3)$$

where, $A_{xx} = \frac{\gamma(H_0 M_0 + M_0^2)}{\alpha_G \omega(2H_0+M_0)}$, $A_{xy} = \frac{M_0}{\alpha_G(2H_0+M_0)}$, the $\gamma$ is the gyromagnetic ratio, $\alpha_G$ is intrinsic Gilbert damping constant, $j$ is the amplitude of the MW current. The components $h_{y'}$ and $h_{z'}$ represent complex numbers with the real ($r$) and imaginary ($i$) parts, which account for their different phases measured with respect to $j(t)$. Finally, Combining with the coordinate transformation applies: $(x', y', z')=(x\cos\theta + y\sin\theta, y\cos\theta - x\sin\theta, z)$, we can separate the nonresonant voltage contributions from $h_x$, $h_y$ and $h_z$:

$$V_0 = V_x j h_x \cdot \sin2\theta\sin\theta + V_y j h_y \cdot \sin2\theta\cos\theta + V_z j h_z \cdot \sin2\theta, \quad (4)$$

where the coefficient $V_x$, $V_y$ and $V_z$ are given by:

$$V_x = \frac{R_A A_{xx} \Delta H}{2M_0 \sqrt{H_0^2 + \Delta H^2}} (\Delta H \sin\varphi_x - H_0 \cos\varphi_x), \quad V_y = \frac{R_A A_{xx} \Delta H}{2M_0 \sqrt{H_0^2 + \Delta H^2}} (H_0 \cos\varphi_y - \Delta H \sin\varphi_y),$$

$$V_z = \frac{-R_A A_{xy} \Delta H}{2M_0 \sqrt{H_0^2 + \Delta H^2}} (H_0 \sin\varphi_z + \Delta H \cos\varphi_z).$$

here the $\varphi_x$, $\varphi_y$ and $\varphi_z$ respectively represent the relative phases between the MW current and the $h$-field in x, y and z directions. From Eq. (4), we find that the three components of $h$-field can be separated through their different angle dependence, i.e., $\sin2\theta\sin\theta$ for $h_x$, $\sin2\theta\cos\theta$ for $h_y$ and $\sin2\theta$ for $h_z$. This enables us to determine the $h$-field vector by fitting the angle dependence of $V_0$ using the $h$-field vectors as the fitting parameters. Besides, the lineshape of the nonresonant voltage, unlike the resonant case, is independent of the magnitude of $H$ since the coefficients $D$ and $L$ show up as a constant. This enables us to measure the angle dependence of $V_0$ without sweeping $H$, which greatly reduces the measurement time. In order to demonstrate the feasibility of the nonresonant method in separating the $h$-field components, we measured the angle dependence of $V_0$ under the in-plane rotating $H$ in the following.

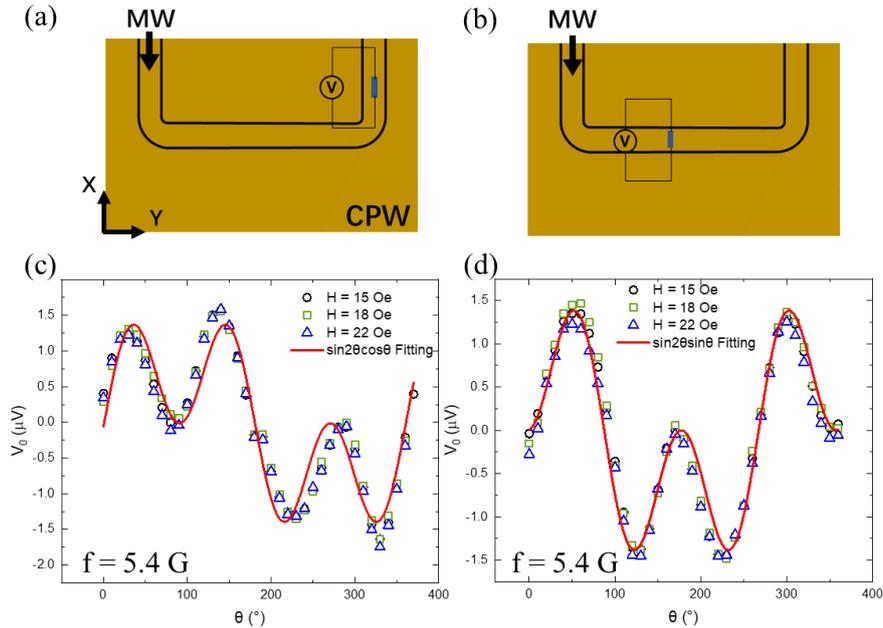

Fig. 2 (a, b) The sketch of the experimental set-up where the Py layer is placed on the signal line in OX and OY direction, respectively. (c, d) The angle dependence of the nonresonant voltage $V_0$ under $f$ = 5.4 GHz and $H$ = 15 Oe, 18 Oe and 22 Oe. The red line is the fitting curve of $\sin2\theta\cos\theta$ and $\sin2\theta\sin\theta$.

The measurement set-ups for the angle dependence are shown in Fig. 2(a) and (b).

The system is analogy to our previous work in mapping the spin Hall angles.[19] The Py stripe is respectively placed along the signal line in OX and OY direction of a coplanar waveguide (CPW) under 5.4 GHz MW excitation, while the DC voltage is always measured along the OX direction, here the applied external magnetic field $H$ around 18 Oe is far less than the resonance field ($H_0$) of Py at 5.4 GHz (~370 Oe). The measured angle dependences of $V_0$ are shown in Fig. 2(c) and (d). All the angle dependences can be well fitted by $\sin2\theta\cos\theta$ and $\sin2\theta\sin\theta$ as indicated by Eq. (4), strongly suggesting that the voltage $V_0$ is induced by the rectification effect. In addition, it is reported that the thermal effect induced by the ANE or SSE can also generate a DC voltage at the nonresonant condition, where the angle dependence between the voltage and $H$ is $\sin\theta$.[20, 21] However, this effect is not obvious in our current measurements as shown by the goodness fitting of the data obtained to Eq. (4). We further checked the field amplitude dependence of $V_0$. The angle dependences of $V_0$ under three different $H$ (15, 18 and 22 Oe) in Fig. 2 (c) and (d) imply that influence of the magnitude of $H$ is negligible for the nonresonant method. As mentioned above, it is sufficient to cause a uniform procession of the magnetization at $H > H_s$ while it is still too weak to be affected by the resonant lineshape at $H \ll H_0$. As a result, the voltage lineshape at $H_s < H \ll H_0$ shows up as a constant independent of the magnitude of $H$.

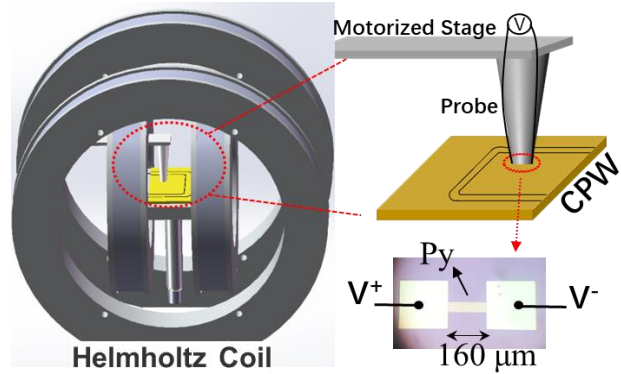

Fig. 3. Schematics of the scanning system. The sensor on the tip of the vertical probe is a patterned Py thin film with the dimensions of 160 μm × 40 μm × 20 nm.

The results of the point-measurement in Fig. 2 can be naturally extended to a spatial map of the $h$-field. The scanning system is schematically shown in Fig. 3, since the $H$ for the nonresonant condition is small, we use a Helmholtz coil as the magnetic field source. In the center of Helmholtz coil, the homogeneous in-plane magnetic field with variable directions can be easily generated by modulating the current in the two coils. The detector of the scanning system is the Py stripe with the dimension of 160 μm×40 μm×20 nm as mentioned above. Note that the detector is of subwavelength size which is far less than the MW wavelength at 5.4 GHz (~5.6 cm). From the AMR and FMR measurements at a shorted microstrip configuration, similar as described in ref. 10, the ratio between the three coefficients in Eq. (4) of the Py stripe is obtained as $V_y = 3.1\ V_x = 17.7\ V_z$ using the parameters: $M_0 = 0.97$ T, $\alpha_G = 0.017$, $R_A = 0.5$ Ω, $\tan\varphi_{x,y,z} = -2.22, 0.33$ and $-0.05$, the measurement of all the parameters is shown in the supplemental materials. In order to spatially map the $h$-field, the Py scanning probe is mounted on a motorized stage, where the distribution of $h$-field can be obtained by the point-by-point scanning above the CPW. The measured distribution of the h-field over the CPW is also compared with the high frequency finite element method (FEM) to demonstrate the feasibility of the scanning method, as described below.

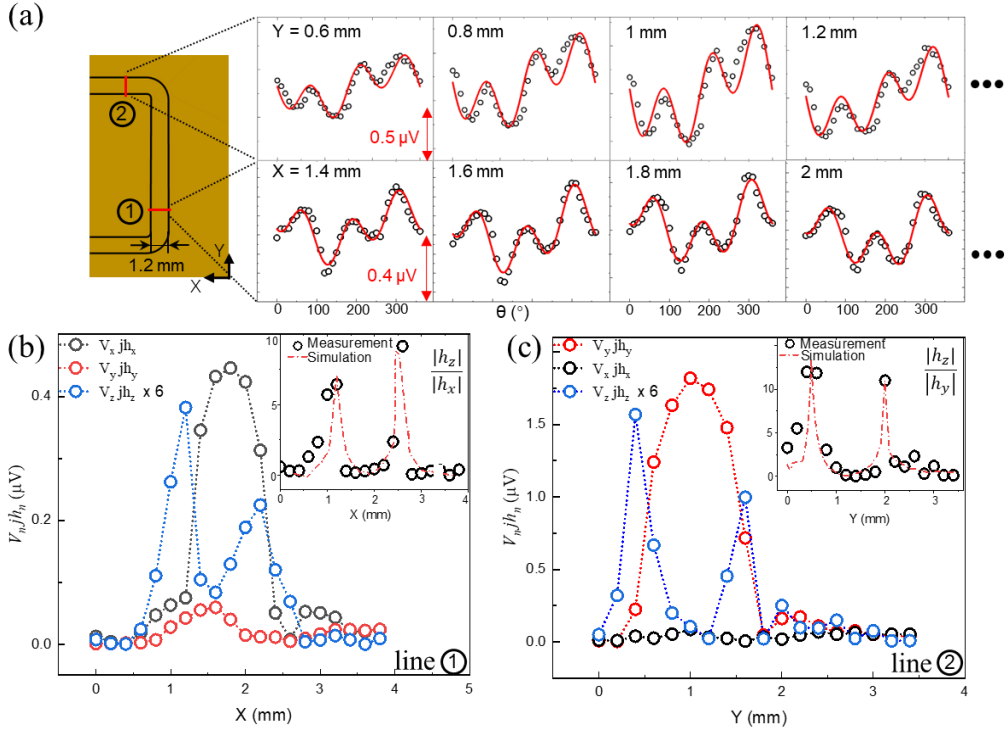

Fig. 4 (a) The sketch of two line-scan measurements and part of the original scan data with fitting. (b, c) The separated coefficients of $\sin2\theta\cos\theta$, $\sin2\theta\sin\theta$ and $\sin2\theta$ of Eq. (4) from the line-scan voltages, the $\sin2\theta$ component is amplified by 6 times for a better comparation. The inset in (b) and (c) are the variation of the measured h-field vector along the scanning line, where the red dashed line represents the simulation results.

We first perform a line-scan over the CPW. Fig. 4(a) shows the sketch of two line-scan experiments across the slits of 3 mm with a step of 0.2 mm. The MW frequency feed into the CPW is also 5.4 GHz and the magnetic field $H$ is 15 Oe. For each point, we increase the field angle $\theta$ from 0° to 360° with a step of 10°, four of the original angle dependence curves are shown Fig. 4(a). All the curves are fitted by Eq. (4) (the red solid lines) to separate the components of h-field. As illustrated by the theory above, the angle dependence reflects the unique feature of the nonresonant rectification effect, where the fitting coefficients are proportional to the amplitude of $jh_x$, $jh_y$ and $jh_z$ respectively. The distributions of h-field vectors along the two scanning lines are directly reflected by the variation of the fitting coefficients as shown in Fig. 4(c) and (d), where the fitting data along with its position is highlighted by dots of three different colors. It can be observed that the MW field is not homogenously distributed above the CPW, especially in the edge of the signal line. This inhomogeneity may be the source of inconsistence among the reported values of the spin Hall coefficient as discussed by Hoffmann and Miao, et al.[12, 22] Combining with the measured value of $V_x$, $V_y$ and $V_z$ mentioned above, we can extract the vector orientation of the h-field in the form of $|h_z|/|h_{x,y}|$ from the two line-scan results, as illustrated by the black dots in the inset of

Fig. 4(b) and (c). Besides, the *h*-field vectors at the same position are also numerically calculated by the FEM method which are illustrated by the red dashed line in the same figure. The good agreement between the measurement and simulation proves the principle of the method works well. By the way, the absolute value of *h*-field vectors can be obtained if the MW current *j* were measured at the same position in principle. In our experiment, the magnitude of *j* is not necessary as the orientation of *h*-field is calculated by the ratio of $jh_x$, $jh_y$ and $jh_z$.

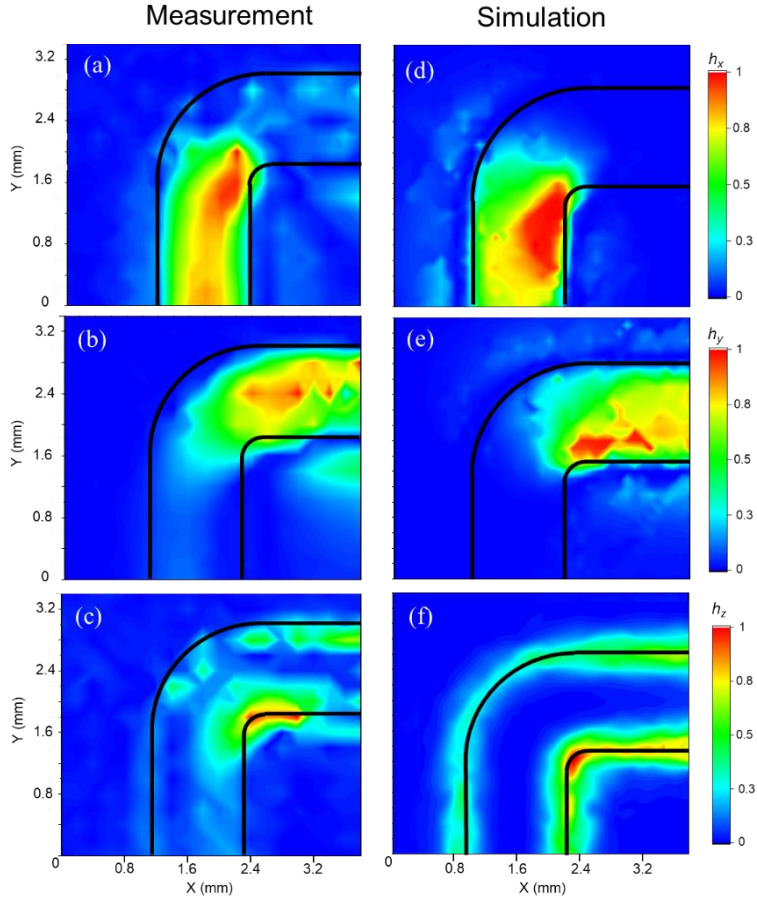

Fig. 5 (a-c) The measured and (d-e) simulated spatial distribution of the *h*-field vector over the CPW. All the images have been normalized with its maximum value.

In the following, we perform a surface scanning above the corner of the CPW where the inhomogeneity is expected to be significant. The right-angled stripline we employ can generate a highly inhomogeneous MW field with a nontrivial field distribution which is linearly polarized. Figure. 5(a-c) show the resulting images of *h*-field vectors with the dimension of 3.8 mm × 3.4 mm, other measurement parameters are the same with the line-scan above. It can be seen that the $h_y$ and $h_x$ components distribute along the signal line in OX and OY direction, while the $h_z$ component mainly focus on the two sides of the stripe. The spatial distribution can be well compared with the numeric results from the simulations by the FEM method, which are shown in Fig. 5(d-f). We believe that the small difference between the simulation and measurement for the *h*-field vectors is caused by the spatial resolution of our imager, since the gap between the signal line and ground line (~ 5 μm) is much smaller than the dimension of the probe (160 μm×40 μm) and the scanning step (0.2 mm).

In conclusion, we have demonstrated that the nonresonant SRE of magnetic thin films can be utilized to detect and image the MW magnetic field vectors on the sub-wavelength scale. Through the measurement of the angle dependence on $H$, we demonstrate the DC voltage at nonresonant condition of a single Py layer originates from the SRE rather than the thermal effect. The mapping results of the CPW are comparable with that from FEM simulations, demonstrating the feasibility of our method. By the way, the resolution can be further increased by reducing the dimension of the magnetic thin film probe. Compared to the conventional SRE method measured at ferromagnetic resonance, the nonresonant approach can operate at a much smaller $H$, which is favorable in system integration. More importantly, the measuring time can be greatly reduced since there is no need to tune the amplitude of $H$ to measure the resonant curves. Our method can be utilized in the MW circuits design and diagnosis, as well as spintronics and quantum information processing where the MW field is involved.